\documentclass[%
reprint,
showpacs,reprintnumbers,
 amsmath,amssymb,
 aps,
prl,
superscriptaddress,
showpacs, showkeys]{revtex4-1}
\usepackage{amsmath}
\usepackage{subfigure}
\usepackage{textgreek}
\usepackage{color}
\usepackage{breqn}
\usepackage{graphicx}
\usepackage{dcolumn}
\usepackage{bm}
\usepackage{hyperref}
\usepackage{textcomp}
\usepackage{xr}
\usepackage{soul}
\hypersetup{bookmarksnumbered, pdfpagemode=UseOutlines, 
colorlinks=true, citecolor=blue, filecolor=blue, linkcolor=blue, urlcolor=blue}

 \newcommand{\beginsupplement}{%
         \setcounter{table}{0}
         \renewcommand{\thetable}{S\arabic{table}}%
         \setcounter{figure}{0}
         \renewcommand{\thefigure}{S\arabic{figure}}
         \setcounter{equation}{0}
         \renewcommand{\theequation}{S\arabic{equation}}%
      }

 \makeatletter
 \let\cat@comma@active\@empty
 \makeatother

\graphicspath{/home/sid/Dropbox/3rd_paper_graphene-TMD
}

\begin{document}


\title{Graphene-WS$_{\text{2}}$ heterostructures for tunable spin injection and spin transport}

\author{S. Omar} 
\thanks{corresponding author}
\email{s.omar@rug.nl}
\affiliation{The Zernike Institute for Advanced Materials University of Groningen Nijenborgh 4 9747 AG, Groningen, The Netherlands}
\author{B.J. van Wees}
\affiliation{The Zernike Institute for Advanced Materials University of Groningen Nijenborgh 4 9747 AG, Groningen, The Netherlands}%
\date{\today}

\begin{abstract}
We report the first measurements of spin injection in to graphene through a 20 nm thick tungsten disulphide (WS$_{\text{2}}$) layer, along with a modified spin relaxation time ($\tau_{\text{s}}$) in graphene in the WS$_{\text{2}}$ environment, via spin-valve and Hanle spin-precession measurements, respectively. First, during the spin-injection into graphene through a WS$_{\text{2}}$-graphene interface, we can tune the interface resistance at different current bias and modify the spin injection efficiency, in a correlation with the conductivity-mismatch theory. Temperature assisted tunneling is identified as a dominant mechanism for the charge transport across the interface.  Second, we measure the spin transport in graphene, underneath the  WS$_{\text{2}}$ crystal and observe  a significant reduction in the $\tau_{\text{s}}$ down to 17 ps in graphene in the WS$_{\text{2}}$ covered region, compared to that in its pristine state. The reduced $\tau_{\text{s}}$ indicates the WS$_{\text{2}}$-proximity induced additional dephasing of the spins in graphene.
\begin{description}
\item[PACS numbers]
 \verb+85.75.-d+, \verb+73.22.Pr+, \verb+75.76.j+
\end{description}
\end{abstract}

\keywords{Spintronics, Graphene, spin injection, tunnel barrier, Schottky barrier, graphene-semiconductor interface, thermally assisted tunneling}
\maketitle

 Graphene, an ideal material for spin transport due to low spin-orbit coupling and small hyperfine interactions \cite{ertler_electron_2009, huertas-hernando_spin-orbit-mediated_2009}, has shown a significant improvement in its spin transport  properties over the years \cite{zomer_long-distance_2012,ingla-aynes_$24-ensuremathmumathrmm$_2015}. However, tuning of the spin injection and transport properties remains illusive for graphene, inhibiting the demonstration of graphene as a spin-transistor \cite{datta_electronic_1990}. There is also a rapidly growing interest in other layered two-dimensional materials such as transition metal dichalcogenides (TMDs) due to their novel properties such as presence of band-gap accompanied by a significant spin orbit coupling up to few hundreds of meV, which is lacking in graphene \cite{2-d-review_2015, Jiang_2012_GW, xu_spin_2014, zhu_giant_2011, latzke_electronic_2015}. In the absence of an inversion center in the lattice, these materials with an odd number of layers also provide access to the novel physical phenomena related to the valley coupled spin degree of freedom of the charge carriers \cite{bawden_spin-valley_2016, yao_valley-dependent_2008, wang_electronics_2012}, which adds extra functionality to these materials. A combination of graphene with these 2-D materials appears to be a plausible option to overcome the aforementioned shortcomings.  
 
 In recent years, there have been a lot of studies on graphene-2-D material heterostructures demonstrating novel charge transport properties across the interface \cite{yu_vertically_2013,georgiou_vertical_2013, myoung_large_2013, lin_barrier_2013}. The 2-D materials such as hexagonal boron nitride (h-BN) can be used in spintronic devices as tunnel barriers for spin injection in graphene, replacing the conventional oxide tunnel barriers  \cite{gurram_spin_2016, kamalakar_enhanced_2014, joiner_graphene-molybdenum_2016}. 
In contrast to an insulating tunnel barrier, the use of band-gap 2-D semiconductors such as TMDs (i.e. MoS$_{\text{2}}$, WS$_{\text{2}}$) during spin-injection, can lead to attractive features such as tuning of the interface resistance along with the induced spin orbit coupling at the graphene-TMD interface  \cite{goswami_spin-orbit_2016, avsar_spinorbit_2014, wei_strong_2016, yang_electrically_2016, wang_origin_2016, tian_electrical_2016, gmitra_trivial_2016}, which in turn can modulate the spin-injection efficiency as well as the spin transport properties in graphene. Recently reported weak localization \cite{wang_strong_2015}, Shubnikov-de Haas magnetoresistance measurements \cite{wang_origin_2016} and spin-Hall experiments \cite{barbaros_graphene-WS2_2014} on graphene-WS$_{\text{2}}$ (Gr-WS$_{\text{2}}$) heterostructures reveal that the spin-lifetime ($\tau_{\text{s}}$)  in graphene is greatly reduced from nanoseconds to picoseconds due to significantly induced spin-orbit coupling ($\sim$ 5-15 meV) in graphene. 
A recent report on spin-transport in Graphene-MoS$_{\text{2}}$ structures demonstrated via spin-valve measurements that the MoS$_{\text{2}}$ flake in the spin transport channel acts as a controllable spin-sink \cite{yan_two-dimensional_2016}.

We report the first measurements of spin injection in to graphene through a 20 nm thick WS$_{\text{2}}$ layer, along with a reduced spin relaxation time in graphene in the WS$_{\text{2}}$ environment, via Hanle spin-precession measurements. By applying a voltage bias between graphene and the semiconducting TMD layer, we tune the interface resistance and modify the spin injection efficiency. We measure a higher spin signal for a higher interface resistance at the injector. In this way spins cannot flow back and get relaxed at the interface or in the bulk WS$_{\text{2}}$, suppressing the contact induced spin relaxation.  We also perform temperature dependent measurements for a Gr-WS$_{\text{2}}$ heterostructure and find that the ideality factor, which is a measure of thermionic emission of the charge carriers across a potential energy barrier, is much greater than one.  It indicates that there are other transport mechanisms such as temperature or field assisted tunneling across the interface, contributing to the  spin injection process in graphene.   
We also measure the spin transport in graphene underneath the WS$_{\text{2}}$ crystal, where we inject and detect the spin accumulation in graphene using the ferromagnetic tunneling contacts. We observe a significant reduction in $\tau_{\text{s}}$, when the spins travel across the WS$_{\text{2}}$ encapsulated region, compared to  $\tau_{\text{s}}$ obtained for the non-encapsulated region. The reduced $\tau_{\text{s}}$  suggests towards an induced spin-orbit coupling/ spin absorption at the Gr-WS$_{\text{2}}$ interface \cite{gmitra_trivial_2016, wang_origin_2016, wang_strong_2015}.

\begin{figure}
 \includegraphics[]{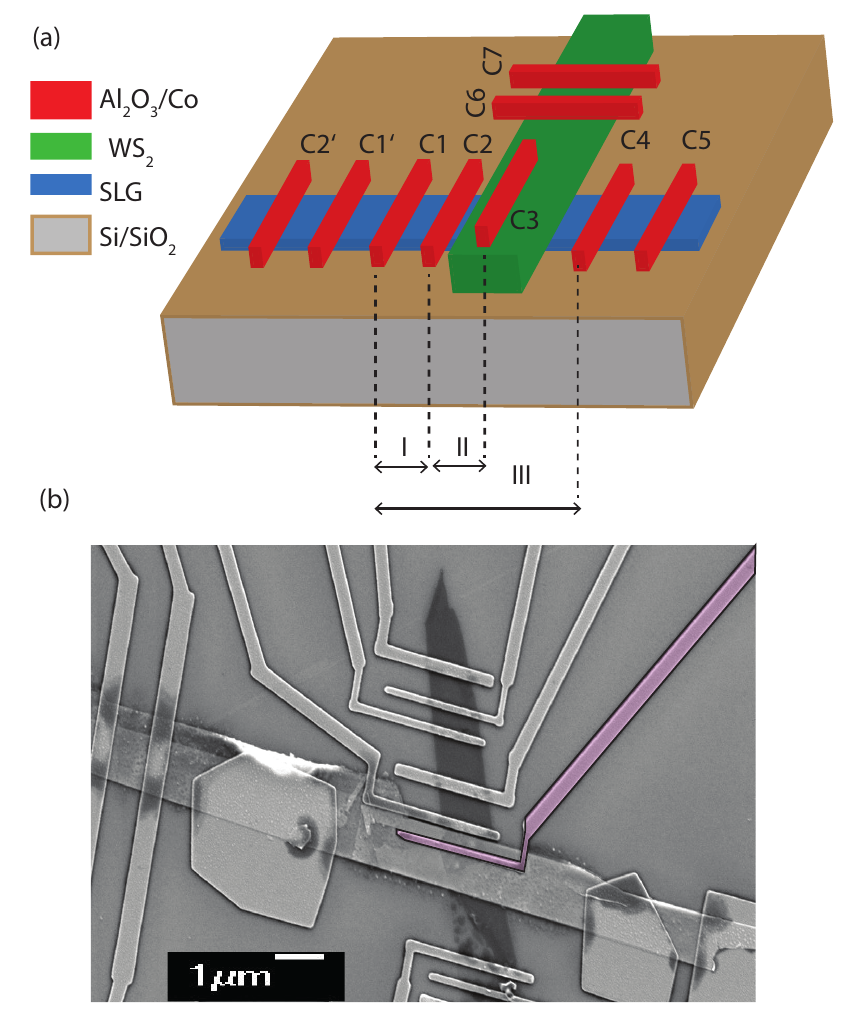}
 \caption{\label{fig:gr-TMD geometry}
  (a)A schematic of graphene-WS$_{\text{2}}$ heterostructure with ferromagnetic contacts with Al$_{\text{2}}$O$_{\text{3}}$ tunnel barriers between cobalt and graphene(WS$_{\text{2}}$). The regions labeled as I, II and III are 2.0 $\mu$m, 2.5 $\mu$m and 6.5 $\mu$m long, respectively. The WS$_{\text{2}}$ covered region is $\sim$ 3.0 $\mu$m.
  (b) A scanning electron microscope (SEM) image of the stack with the ferromagnetic contacts. The pink electrode is used as a spin injector into graphene through WS$_{\text{2}}$.}
 \end{figure}
 
We prepare graphene-WS$_{\text{2}}$ heterostructures via a pick up transfer method, described in ref.~\cite{zomer_fast_2014} (see supplementary for details) using a single layer graphene and a thick WS$_{\text{2}}$ flake, obtained via scotch tape exfoliation procedure. This dry transfer method enables the formation of a clean and chemical free Gr-WS$_{\text{2}}$ interface, which has been reported to have less impurities and superior charge transport properties \cite{moriya_large_2014} compared to the stacks prepared via CVD grown 2-D materials \cite{yu_vertically_2013}. Moreover, we transfer a thick WS$_{\text{2}}$ flake on to graphene, in order to reduce the bubble formation during the transfer process \cite{yang_tunable_2016}. The ferromagnetic (FM) contacts are patterned both on graphene and on WS$_{\text{2}}$ via electron beam lithography on the PMMA (poly (methyl methacrylate)) coated Gr-WS$_{\text{2}}$ stack. Then, 0.6 nm of aluminum (Al) is deposited in two steps, each step of 0.3 nm of Al deposition followed by in-situ oxidation by pure O$_{\text{2}}$ to form an oxide tunnel barrier to overcome the conductivity mismatch problem \cite{maassen_contacts}. On top of the oxide barrier we deposit 55 nm of cobalt for the spin polarized contacts. To prevent the oxidation of the ferromagnetic electrodes, the contacts are covered with 3 nm thick aluminum layer.


We characterize the charge and spin transport in graphene at three different regimes, labeled as I, II and III in Fig.~\ref{fig:gr-TMD geometry}(a): i) non-encapsulated region (I) as a reference; ii) through the WS$_{\text{2}}$ crystal (II), where  the charge/spin current is injected from a ferromagnet C3 on top of the TMD crystal in to graphene and is detected in graphene (Fig.~\ref{fig:gr-TMD geometry}(b)) \footnote{A SEM image was recorded after the measurements. The graphene and WS$_{\text{2}}$ flake were damaged after the measurements during storage.}, iii) across the encapsulated region ($\sim$ 3 $\mu$m) (III), where a charge/spin current is injected in graphene on one side of the TMD  via contact C2 and is detected on the other side via contact C4.

\begin{figure}
 \includegraphics[]{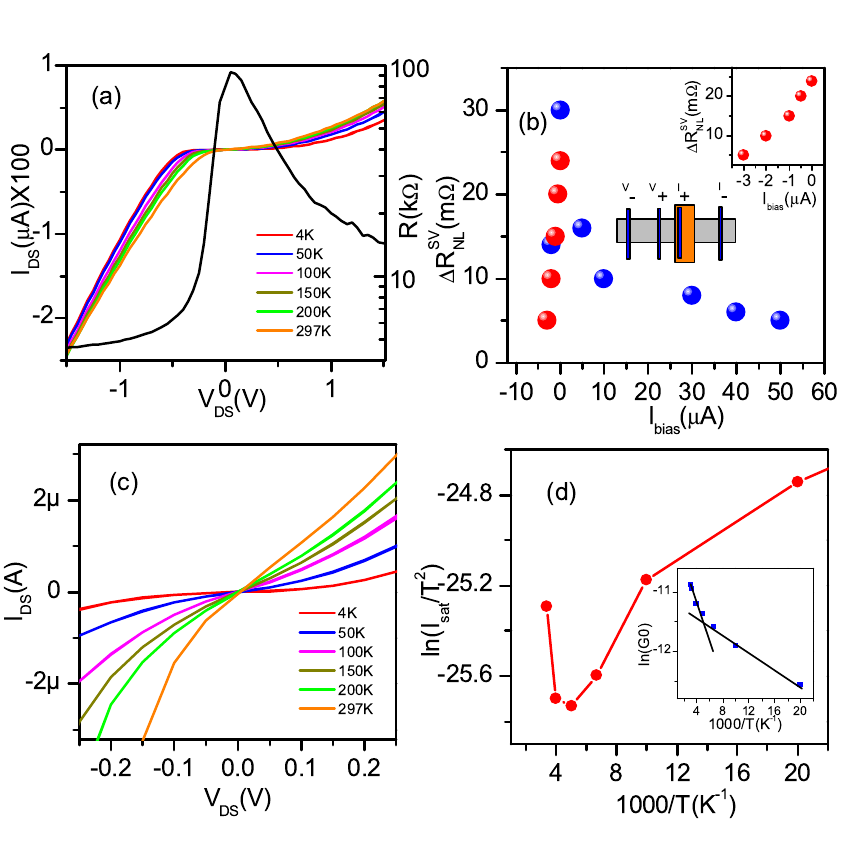}
 \caption{\label{fig:hetero IV}
 (a) 2-probe I-V measurements for the graphene-WS$_{\text{2}}$ interface at different temperatures. Here WS$_{\text{2}}$ is the positive electrode. On the right y-axis, the interface resistance  (d$V_{\text{DS}}$/d$I_{\text{DS}}$) is plotted as a function of $V_{\text{DS}}$ for room temperature shows a non-linear behavior. (b) A plot of non-local spin signal ($\triangle R_{\text{NL}}^{\text{SV}}= R_{\text{P}}-R_{\text{AP}}$) as a function of dc bias current, injected through WS$_{\text{2}}$. The measurement scheme is shown in the cartoon. The current is applied between contacts C3-C5 of Fig.~\ref{fig:gr-TMD geometry}(a) and a voltage is measured between contacts C2-C2\textasciiacute. The spin signal is significantly reduced for a lower interface resistance. A zoomed in version for the negative $I_{\text{bias}}$-spin signal dependence is shown in the inset. (c) Zoomed in $I_{\text{DS}}-V_{\text{DS}}$ plot around zero bias  shows more symmetric behavior, suggesting the dominance of tunnel transport over thermionic emission. (d) ln (I$_{\text{DS}}$) versus 1000/T plot shows a cross-over temperature ($T_{\text{cross}}$) around 250K from negative to positive slope, which is a signature of cross over from thermionic emission to thermal assisted tunneling in a broad temperature range. In the inset, the slope of logarithmic conductance changes around the same $T_{\text{cross}}$, implying the same cross-over of conduction mechanisms.  \label{fig:bias dep}}     
\end{figure}

All the measurements are performed using a cryostat in vacuum  ($\sim$ 1 $\times$ 10$^{-7}$ mbar) at different temperatures  between 4K and 297K. The graphene resistivity was characterized via lock-in detection (f=27.7 Hz) using a four probe method by applying a current between contacts C2-C2\textasciiacute and measuring a voltage drop between C1-C1\textasciiacute, in order to eliminate the effect of the contacts. The graphene sheet resistance ($R_{\text{sq}}$) for the non-encapsulated region (region I) is $\sim$ 400 $\Omega$  (charge carrier density $\sim$ 10$^{\text{13}} cm^{\text{-2}}$ \footnote{Since we do not have a working back gate for the reported sample, we estimate the carrier density $n$ from the relation $n=1/R_{\text{sq}}e\mu$, where $e$ is the electron charge and $\mu$ is the field effect mobility. Here, we take $\mu \sim$ 5000 $cm^{\text{2}}V^{\text{-1}}s^{\text{-1}}$ which is usually obtained for good quality graphene samples on SiO$_{2}$ substrate} where )  i.e. three times lower than for region III ($\sim$ 1.2 k$\Omega$), indicating that graphene is less doped underneath the WS$_{\text{2}}$ crystal. The contact resistances for the FM electrodes were characterized using a three probe connection scheme, where an ac current is applied between contacts C1-C2 and a voltage drop is measured between C1-C2\textasciiacute. For the FM contacts on graphene, we measure a very low contact resistance ($R_{\text{C}}^{\text{FM}} \sim$ 200 $\Omega$), putting our contacts in the so called conductivity mismatched regime  where the contacts influence the spin transport properties of graphene \cite{maassen_contacts}. Since we also fabricate the contacts  on the WS$_{\text{2}}$ flake out side the WS$_{\text{2}}$-graphene interface (i.e. contacts C6-C7 on WS$_{\text{2}}$ in Fig.~\ref{fig:gr-TMD geometry}(a)), the channel resistance ($R_{\text{sq}}^{WS_{2}} \sim$ 70 k$\Omega$) and the contact resistance of the FM electrodes on the WS$_{\text{2}}$ flake ($\sim$ 2-3 k$\Omega$)  can be characterized independently. The I-V behavior of the FM contacts, both on graphene and on WS$_{\text{2}}$ is characterized independently using a Keithley 2410 dc source meter. The measured I-V behavior has a linear dependence at low bias, which becomes slightly non-linear at higher bias values ($R_{\text{c}}\leq$4k$\Omega$) (see supplementary). Next, we measure a non-linear I-V behavior across the graphene-TMD interface via FM contacts (Fig.~\ref{fig:hetero IV}(a)) which is clearly is different from the I-V behavior at the WS$_{\text{2}}$-AlO$_{\text{2}}$-Co interface, and is dominated by the Gr-WS$_{\text{2}}$ interface. The observed non-linearity can be easily attributed to the presence of a potential energy barrier present only at the Gr-WS$_{\text{2}}$ interface \cite{moriya_large_2014, yamaguchi_tunneling_2014, tian_electrical_2016}. 

\begin{figure}
 \includegraphics{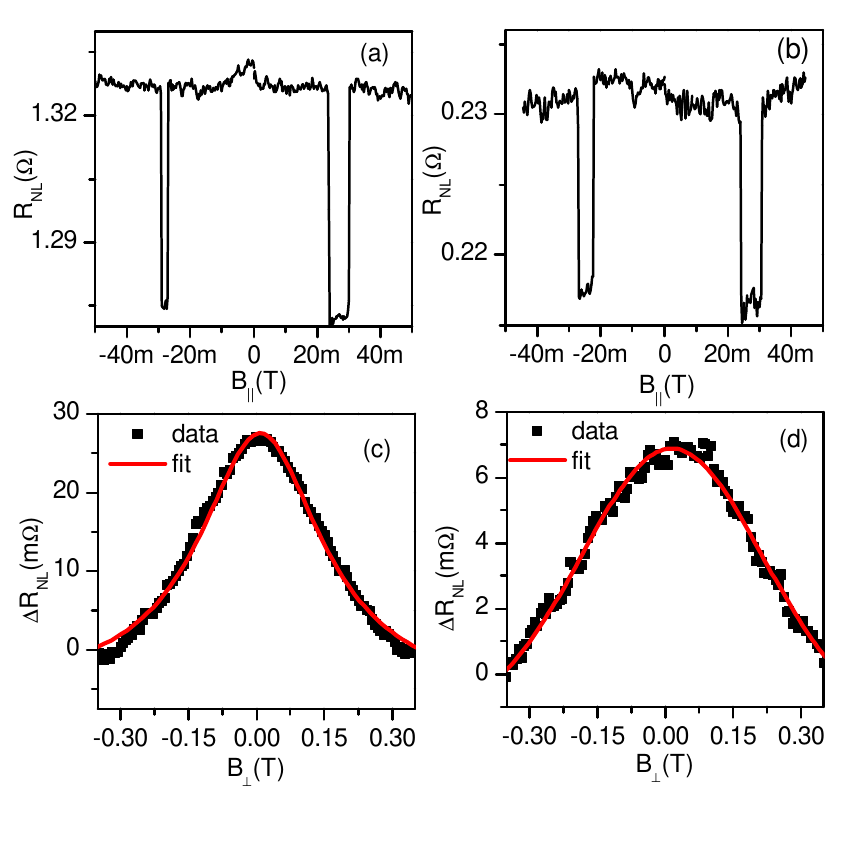}
  \caption{\label{fig:spin valve hanle non-encap}
 (a) Spin-valve and (c) Hanle precession measurements for non-encapsulated part (region I), with the FM contacts used for spin transport. (b) spin-valve and (d) Hanle measurements for region II, where WS$_{\text{2}}$, inserted between FM contacts and graphene is used as spin-injector and the FM  contact on graphene is used for measuring the spin accumulation.} 
  \end{figure}

Spin transport is measured in a four probe non-local detection scheme i.e. in region I, by applying a current between contacts C1 and C2\textasciiacute and measuring the voltage associated with the spin-accumulation between contacts C2-C5. This method decouples the paths of the spin and charge transport and thus minimizes the contribution of the charge signal to the measured spin signal \cite{Tombros_nature}. For the spin valve measurements, we first apply an in-plane high magnetic field ($B_{\parallel}$) along the easy axes of the ferromagnets to align their relative magnetization. Then, the magnetic field is swept in the opposite direction to reverse the magnetization of the ferromagnets selectively based on their coercivity. We measure a sharp transition in the non-local signal ($R_{\text{NL}}=V_{\text{NL}}/I_{\text{ac}}$) when the individual electrodes switch their magnetization direction  (Fig.~\ref{fig:spin valve hanle non-encap}(a)). For the Hanle precession measurements, an out of plane magnetic field ($B_{\perp}$) is applied to precess the injected spins around the applied field for a fixed magnetization configuration of the ferromagnetic electrodes. While diffusing the spins precess around the applied magnetic field and dephase, showing a decrease in the magnitude of measured spin accumulation for higher fields.

\begin{figure}
 \includegraphics[]{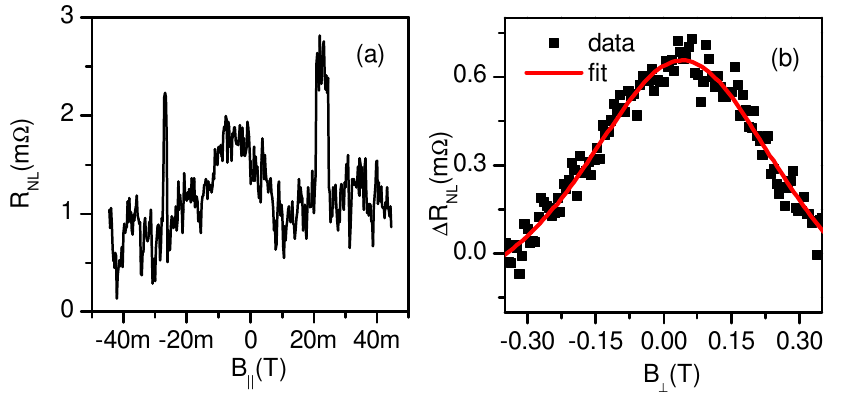}
\caption{(a) Spin valve and (b) Hanle measurement for the encapsulated part (region III), where a spin current is injected via a FM contact C2 (Fig.~\ref{fig:gr-TMD geometry}(a)) on one side of the WS$_{\text{2}}$ crystal and is measured via contact C4 on the other side after  traveling underneath the WS$_{\text{2}}$ environment. \label{fig:spin valve hanle encap}}
 \end{figure}

 
 
 Hanle precession measurements for region I, II and III are shown in Fig.~\ref{fig:spin valve hanle non-encap} and Fig.~\ref{fig:spin valve hanle encap}, respectively. With these measurements, we fit the Hanle signal $\triangle R_{\text{NL}}=(R_{\text{P}}-R_{\text{AP}})/2$, where $R_{\text{P(AP)}}$ is the Hanle signal measured for the parallel (anti-parallel) magnetization of the the injector-detector pair.  We extract the spin diffusion coefficient $D_{\text{s}}$ and spin relaxation time $\tau_{\text{s}}$, following the procedure described in ref.~\cite{Tombros_nature} and use them to calculate the contact polarization ($P$). For region I, we obtain $D_{\text{s}}\sim $0.09 m$^2$/s, $\tau_{\text{s}} \sim$ 40 ps and $P \sim$ 2.3\%, for region II, $D_{\text{s}} \sim$ 0.09 m$^2$/s, $\tau_{\text{s}} \sim$ 17 ps and $P \sim$ 2.7\% and for region III, we observe a  $D_{\text{s}} \sim$ 0.1-0.5 m$^2$/s, $\tau_{\text{s}} \sim$ 18 ps and $P \sim$ 0.3\%. For region III, the $D_{\text{s}}$ value obtained via Hanle fitting, is sensitive to the selection of the spin-independent background, which cannot be uniquely determined here. However, we consistently obtain a reduced $\tau_{\text{s}}$ for region III $\sim$ 17 ps-20 ps. In conclusion, a reduced $\tau_{\text{s}} \leq$ 20 ps is obtained via Hanle measurements for both region II and III where the spin-transport occurs underneath the WS$_{\text{2}}$ flake. Here, we would like to emphasize that a low $\tau_{\text{s}}$ observed in region I is probably due to the contact induced spin relaxation. For a lower contact resistance the spins can easily flow back to the contact and get relaxed, and one would obtain a lower $\tau_{\text{s}}$.  The actual value for $\tau_{\text{s}}$ can be estimated by following the procedure as described by Maassen \textit{et al.} \cite{maassen_contacts}. In this method, the effect of the contact induced spin relaxation can be quantified via \textasciigrave $ R$\textasciiacute parameter ($R=R_{\text{c}}\lambda_{\text{s}}/R_{\text{s}}$), which is a ratio of the contact resistance $R_{\text{c}}$ to the spin resistance ($R_{\text{s}}$) of the transport channel i.e. $R_{\text{s}}=R_{\text{sq}} \lambda_{\text{s}}/W$. Here $\lambda_{\text{s}}$ is the spin relaxation length in graphene, and $W$ is the width of the graphene flake. Based on this method, the actual $\tau_{\text{s}}$  will be three times higher than the fitted $\tau_{\text{s}}$ for the non-encapsulated region. For region II and III, the correction factor is $\sim$ 1, implying that the $\tau_{\text{s}}$ obtained via Hanle fitting procedure represents the true spin relaxation time, confirming that the spin transport is affected significantly underneath the high spin-orbit coupled material i.e. WS$_{\text{2}}$ in our case. 

 In order to determine the nature of charge and spin transport through the  Gr-WS$_{\text{2}}$ interface, we characterize the interface resistance as a function of applied bias and temperature (Fig.~\ref{fig:hetero IV}(a)). There is an increase in the zero bias interface resistance from 80 k$\Omega$ -500 k$\Omega$ from RT to 4K. A non-linear current-voltage characteristics of the Gr-WS$_{\text{2}}$ heterostructure indicates a potential energy barrier formation at the interface. We analyze the interface I-V characteristics with the standard Schottky-Mott model at higher bias ($V_{\text{DS}} >$ 3$k_{\text{B}}T/e$), which is described by the following equation:
 
 \begin{equation}
  I_{\text{DS}}= AA^*T^2 \exp{\frac{eV_{\text{DS}}}{nk_{\text{B}}T}} \exp{\frac{-e\phi_{\text{B}}}{k_{\text{B}}T}}
  \label{Schottky}
 \end{equation}
 
 where $I_{\text{DS}}$ is the current flowing through Gr-WS$_{\text{2}}$ interface, A is the area of the interface,  $V_{\text{DS}}$ is the voltage drop across the interface, $n$ is the ideality factor ($n$=1, for ideal Schottky diode), $A^*$ is the Richardson constant, $\phi_{\text{B}}$ is the effective Schottky barrier height at the Gr-WS$_{\text{2}}$ interface, $e$ is the electronic charge, $k_{\text{B}}$ is the Boltzmann constant and $T$ is the temperature of the device. 
 We extract the ideality factor $n$ and the saturation current $I_{\text{0}}=AA^*T^2\exp{\frac{-e\phi_{\text{B}}}{k_{\text{B}}T}} $ for different temperatures from the slope and the intercept of ln($I_{\text{DS}}$)-$V_{\text{DS}}$ curve of Eq.~\ref{Schottky}. The ideality factor $n$ is highly temperature dependent ($\geq$ 3), exhibiting a strong deviation from  thermionic emission theory at lower temperatures.  In Fig.~\ref{fig:hetero IV}(d), ln ($I_{\text{0}}/T^{\text{2}}$) versus 1000/T plot shows a positive slope for the major temperature range and the slope is only negative for very high temperatures, indicating that the electron transport mechanism in our sample is dominated by thermally assisted tunneling or field assisted tunneling, not by the thermionic emission.  
We also try to fit the zero-bias conductance-temperature dependence (inset Fig.~\ref{fig:hetero IV} (d)) with the tunnel transport model, described in ref.\cite{yamaguchi_tunneling_2014}. A plot of temperature-zero-bias conductivity shows the crossover of the two mechanisms around 200 K around the same temperature as reported in the ref.~\cite{georgiou_vertical_2013}. However, a reasonable agreement between the data and the fit is not obtained, possibly because of the moderate temperature dependence of the conductivity at the lower temperature.  All the presented analyses support temperature assisted tunneling as a dominant charge transport mechanism across the Gr-WS$_{\text{2}}$ interface. 

Novelty of the Gr-WS$_{\text{2}}$ interface is its bias dependent resistance, which in combination with the FM contact can be used to modify the conductivity mismatch condition for the spin injection. Since the charge transport across the interface is dominated by temperature assisted tunneling, it is possible to inject spins in to graphene through TMD due to small Schottky barrier and low depletion width at the interface, in contrast to the case of Si spin-valves, where no spin transport is measured due to a higher Schottky barrier at the ferromagnet-silicon interface of the injector and the detector electrodes \cite{jansen_detection_2007, jansen_electrical_2010}. In order to bias the interface, we use a dc current source at different currents with a fixed small ac current (0.5 $\mu$A) superimposed on it. The spin accumulation is measured non-locally via the lock-in detection (see supplementary for bias dependent spin-valve measurements). On changing the bias current, the interface resistance is modulated from 15 k$\Omega$ -100 k$\Omega$ for the positive current bias and to 4 k$\Omega$ for the negative bias. The change in the resistance is much sharper for the negative bias regime. The spin signal decreases slowly from 50 m$\Omega$-to 5 m$\Omega$ for the positive current range of 50 $\mu$A. For the negative bias, we observe similar change in within 5 $\mu$A current range, suggesting a strong correlation between the interface resistance and spin injection efficiency. However, for the oxide tunnel barriers with 4 k$\Omega$ interface resistance, one can get a reasonable spin injection, as this situation is not a poor conductivity mismatch. Even for smaller interface resistance in region I, we were able to achieve spin injection. A negligible spin injection for a reasonable interface resistance, suggests towards either low spin polarization of the interface as it is not a pure tunneling or additional spin-relaxation underneath the WS$_{\text{2}}$ encapsulated region. Also, we were unable to detect any spin accumulation on the other side of WS$_{\text{2}}$ (contact C4), probably due to the combined effect of poor spin injection through the TMD, accompanied by relatively long distance experienced by spins underneath WS$_{\text{2}}$, before getting detected.  However, from the present measurements, we cannot comment on the the source of enhanced spin relaxation process.

In conclusion, we demonstrate that spin transport across the WS$_{\text{2}}$ encapsulated region and the spin transport is reduced in the proximity of the TMD crystal. At present, we cannot comment on the source of additional spin-relaxation whether it is caused by proximity induced spin orbit coupling or due to spin absorption via the TMD. We also demonstrate the use of WS$_{\text{2}}$ as a bias dependent spin injector due to the non-linear charge transport properties of Gr-WS$_{\text{2}}$ interface. Via a temperature-dependent charge transport analysis, we find out that the dominating charge transport mechanism across the interface is thermally assisted tunneling, which helps the spins to tunnel through the Schottky barrier \footnote{During the submission process, we became aware of a similar work \cite{dankert_all-electrical_2016}, where a reduced $\tau_{\text{s}}$ was obtained via Hanle measurements underneath the MoS$_{\text{2}}$ encapsulated region, similar to the results we obtain.}. 
 
We acknowledge  J. G. Holstein, H. M. de Roosz and H. Adema for their technical assistance. This research work was financed under EU-graphene flagship program (637100) and supported by the Zernike Institute for Advanced Materials and the Netherlands Organization for Scientific Research (NWO).


%

\newpage
\newpage\null\thispagestyle{empty}\newpage

\beginsupplement
\begin{center}
 \textbf{\large Supplementary Information}
\end{center}

\begin{figure}[!ht]
 \includegraphics{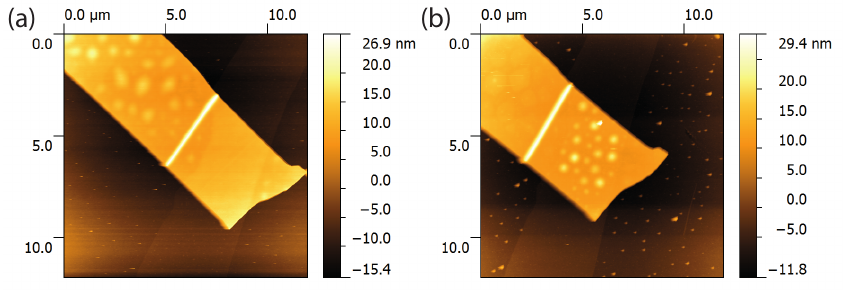}
  \caption{\label{stack 4}
 A graphene-WS$_{\text{2}}$ stack (a) before annealing and after (b) annealing. For the thin WS$_{\text{2}}$ flakes ($t_{WS_2} \sim$ 5nm), the graphene flake underneath the WS$_{\text{2}}$ flake can be seen through. Here, the annealing step does not seem to improve the interface quality significantly.} 
  \end{figure}

Due to the layered structure of graphene and WS$_{\text{2}}$, these materials can be exfoliated via a scotch-tape method. The graphene flake is exfoliated from a bulk HOPG (highly oriented pyrolytic graphite) ZYA grade crystal (supplier: SPI) on to a pre-cleaned Si-SiO$_{\text{2}}$ substrate ($t_{\text{SiO}_{\text{2}}}$=300 nm). A single layer is identified via an optical microscope.  A WS$_{\text{2}}$ flake (spplier: HQ Graphene) is exfoliated on a vsicoelastic PDMS (polydimethylsiloxane) substrate. The freshly cleaved WS$_{\text{2}}$ flake on the PDMS is brought in contact with the graphene flake  in a transfer stage. Since the adhesion of WS$_{\text{2}}$ on the PDMS stamp is relatively weak compared to the GrW van der Waals interaction and the SiO$_{\text{2}}$-WS$_{\text{2}}$ adhesion, the flake is easily stacked on to the desired graphene flake. As a result a smooth interface is formed, which is identified with an atomic force microscope (AFM). For a thick WS$_{\text{2}}$ flake the bubble formation during the transfer procedure is less likely and in the AFM we do not observe any bubbles at the graphene-WS$_{\text{2}}$ interface. The prepared stacks are  annealed at 250$^{\circ}$C for 5 hours in an Ar-H$_{\text{2}}$ environment for removing the residue polymers. However, for most of the stacks, we only observe that the uncovered graphene and WS$_{\text{2}}$ flake looks more clean after the annealing steps, leaving the graphene-WS$_{\text{2}}$ interface more or less similar to that before the annealing (Fig.~\ref{stack 4}).

 \begin{figure}[]
 \includegraphics{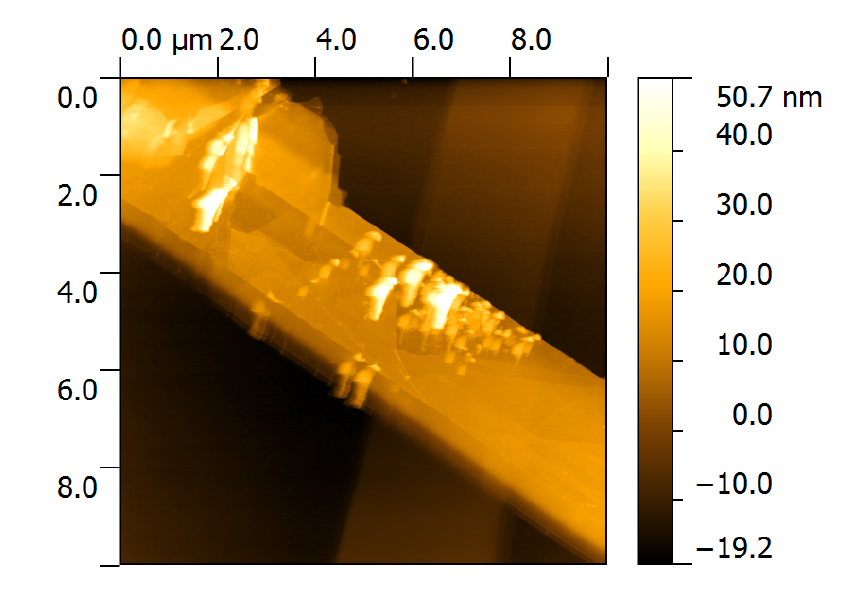}
  \caption{\label{stack 7}
 A graphene-WS$_{\text{2}}$ stack formed with thick WS$_{\text{2}}$ flakes ($t_{WS_2} \sim$ 20nm)} 
  \end{figure}
  
 For the stack with a thick WS$_{\text{2}}\sim$ 20 nm flake, transferred on to a graphene flake, we do not see a bubble formation at the graphene-WS$_{\text{2}}$ interface (Fig.~\ref{stack 7}). Still the furnace annealing is performed to clean the polymer residues from the uncovered surfaces.
 
 \section{Contact I-V for AlO$_x$/Co contacts}

\begin{figure}[]
 \includegraphics{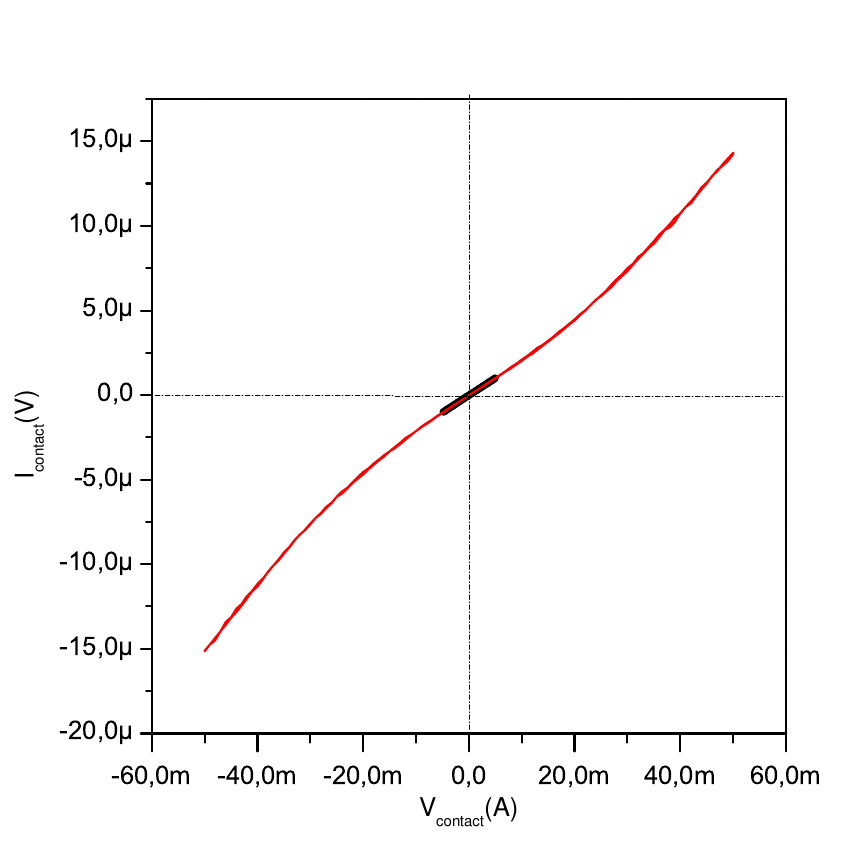}
  \caption{\label{contact WS2}
 I-V characteristics of a Co/AlO$_{\text{x}}$ contact on the WS$_{\text{2}}$ flake.} 
  \end{figure}

    We characterize the contact resistance via three probe connection scheme where a voltage drop across the contact is measured while a current flowing through it. Since we also deposit an insulating oxide layer between the transport channel (graphene or WS$_{\text{2}}$) and the ferromagnet, the contact resistance can deviate from its ohmic behavior depending on the current/voltage bias applied on it. For this purpose, a dc  source meter (KTH 2410) is used. For the FM contacts on WS$_{\text{2}}$, we see a linear  I-V behavior for WS$_{\text{2}}$ in a low bias regime which becomes slightly non-linear at high bias values (Fig.~\ref{contact WS2}). Still, the observed non-linearity is not as strong as measured for Gr-WS$_{\text{2}}$ interface.
    
    \section{Bias dependent Spin valve measurements}
    
\begin{figure}[]
 \includegraphics{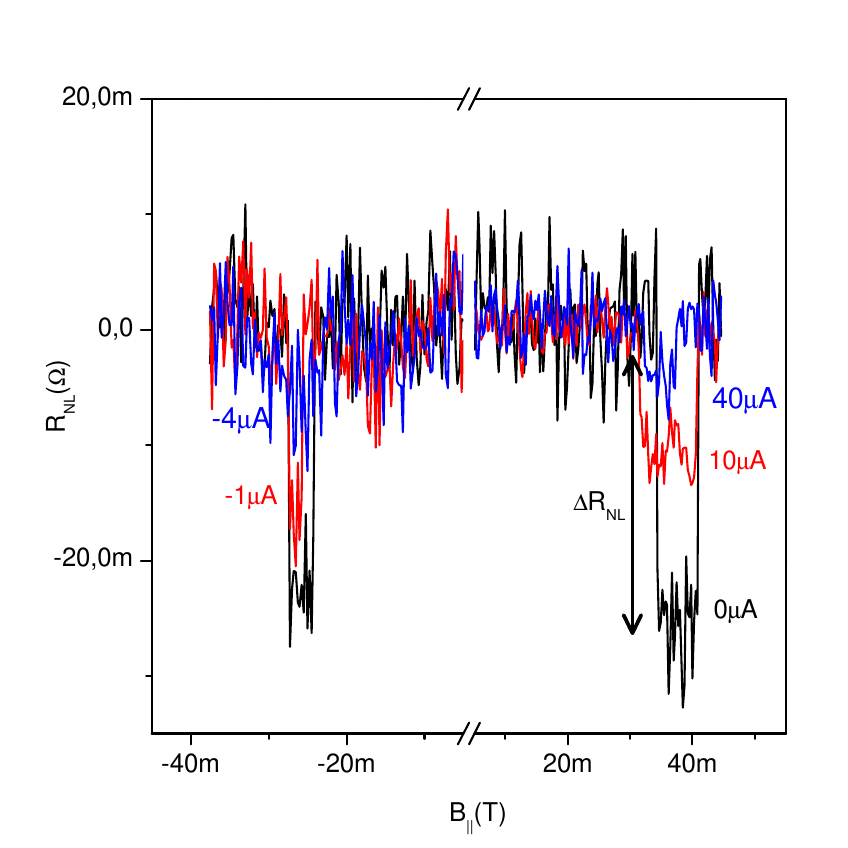}
  \caption{\label{bias pn}
 Spin valve measurements as a function of dc current bias. Spin vale measurements for different positivie (negative) dc bias values are plotted on the right (left) half of the plot ($B_{||}>$ 0T ($<$ 0T)). } 
  \end{figure}

    We perform bias dependent spin-valve measurements by applying a finite dc current (-10 $\mu$A- 50$\mu$A) along with a fixed ac current (= 0.5 $\mu$A) and measure the spin accumulation non-locally via lock-in detection method. By applying different dc bias, we modify the interface resistance and the spin-injection efficiency of the injector electrode. At a high interface resistance of the injector, we measure a higher spin accumulation due to improved spin-injection efficiency. On reducing the interface resistance as a function of positive(negative) bias, the spin signal is reduced due to a lower spin-injection efficiency(Fig.~\ref{bias pn}). 

\end{document}